\documentclass[preprint2,twoside]{hwo}

\usepackage{lipsum}

\input{hwo.h}

\begin{document}

\title{\textbf{\LARGE Habitable Worlds Observatory:\ The Nature of the First Stars}}

\author {\textbf{\large Ian U.\ Roederer,$^{1}$ Rana Ezzeddine,$^2$ Jennifer S.\ Sobeck,$^3$ }}

\affil{$^1$\small\it Department of Physics, North Carolina State University, Raleigh, North Carolina, USA; \email{iuroederer@ncsu.edu}}

\affil{$^2$\small\it Department of Astronomy, University of Florida, Gainesville, Florida, USA}

\affil{$^3$\small\it Caltech/IPAC, Pasadena, California, USA}


\author{\footnotesize{\bf Endorsed by:}
Borja Anguiano (CEFCA),
Narsireddy Anugu (Georgia State University),
David Barckhoff (University of Pittsburgh),
St\'{e}phane Blondin (Aix Marseille Univ/CNRS/LAM),
Howard Bond (Penn State University),
Luca Casagrande (Australian National University),
Annalisa De Cia (European Southern Observatory),
Annalisa Citro (University of Minnesota),
Melanie Crowson (American Public University),
Jose M.\ Diego (Instituto de Fisica de Cantabria),
Emma Friedman (NASA GSFC),
Lukas Furtak (Ben-Gurion University of the Negev),
Farhanul Hasan (Space Telescope Science Institute),
Natalie Hinkel (Louisiana State University),
Joris Josiek (ZAH/ARI, Universit\"{a}t Heidelberg),
Pierre Kervella (Paris Observatory \& CNRS IRL FCLA),
Ji\v{r}\'{i} Krti\v{c}ka (Masaryk University),
Ariane Lan\c{c}on (Observatoire astronomique de Strasbourg - France),
Alex Lazarian (UW-Madison),
Eunjeong Lee (EisKosmos (CROASAEN), Inc.),
Valentina D'Odorico (INAF Trieste),
Julia Roman-Duval (Space Telescope Science Institute),
Shivani Shah (North Carolina State University),
Josh Simon (Carnegie Observatories),
Melinda Soares-Furtado (UW-Madison),
Frank Soboczenski (University of York \& King's College London),
Heloise Stevance (University of Oxford),
David Traore (ORBIT),
Andrew Wetzel (University of California, Davis),
John Wise (Georgia Institute of Technology)
}



\begin{abstract}
We present the science case for characterizing the nature of the first stars 
using the Habitable Worlds Observatory (HWO).~
High-resolution ultraviolet (UV) spectroscopy with the HWO
has the potential to confirm any surviving low-mass 
zero-metallicity first stars by placing 
unprecedented low limits on their metal abundances.
It also has the potential to substantially increase the number of elements
detectable in the spectra of 
known long-lived low-mass stars,
which exhibit extremely low metal abundances that
reveal the metals produced by the first stars.
Elements important for this science case
with UV transitions include
C, Mg, Al, Si, P, S, Sc, V, Cr, Mn, Fe, Co, Ni, Cu, and Zn.
HWO would expand the discovery space when compared with
the Hubble Space Telescope by enabling 
high-resolution UV spectroscopy
for much fainter stars throughout the Milky Way
and neighboring stellar systems.
  \\
  \\
\end{abstract}

\vspace{2cm}

\section{Science Goal}

What was the nature of the first stars?
The first stars in the universe formed 
from the clouds of primordial hydrogen and helium 
created shortly after the Big Bang. 
These first stars, also known as Population~III (Pop~III), 
are expected to have been massive and short lived.
They produced the first metals through fusion reactions 
or during the supernova explosions that ended their lives.
These first metals seeded the interstellar medium 
and enabled the first low-mass long-lived stars to form. 
No metal-free ``first stars'' have yet been found, 
so their nature and end states remain unknown observationally. 

These stars connect the primordial hydrogen and helium to stars today, 
including our Sun. 
Understanding the nature of the first stars is a key part 
of addressing two of the broad themes identified by the 
Astro 2020 Decadal Survey \citep{astro2020}: 
``New Messengers and New Physics”'' and ``Cosmic Ecosystems.'' 
Only the Habitable Worlds Observatory (HWO) 
has the high spectral resolution in the ultraviolet (UV) 
necessary to detect the elements that distinguish the first stars 
and characterize their nature.

\section{Science Objective}

First-generation metal-free Pop III stars 
would have formed in the Milky Way and its surrounding stellar systems
\citep{frebel15araa}.
The elements in each of these stars 
reflect the elements in the gas from which each star formed. 
The elemental abundances in the star observed today 
provide the observational constraint on the supernova yields 
predicted by models \citep{tominaga14,salvadori19},
and the physics of the supernovae themselves \citep{koutsouridou23}.
The objective is twofold.

\begin{figure*}[ht]
\begin{center}
\includegraphics[width=0.7\textwidth]{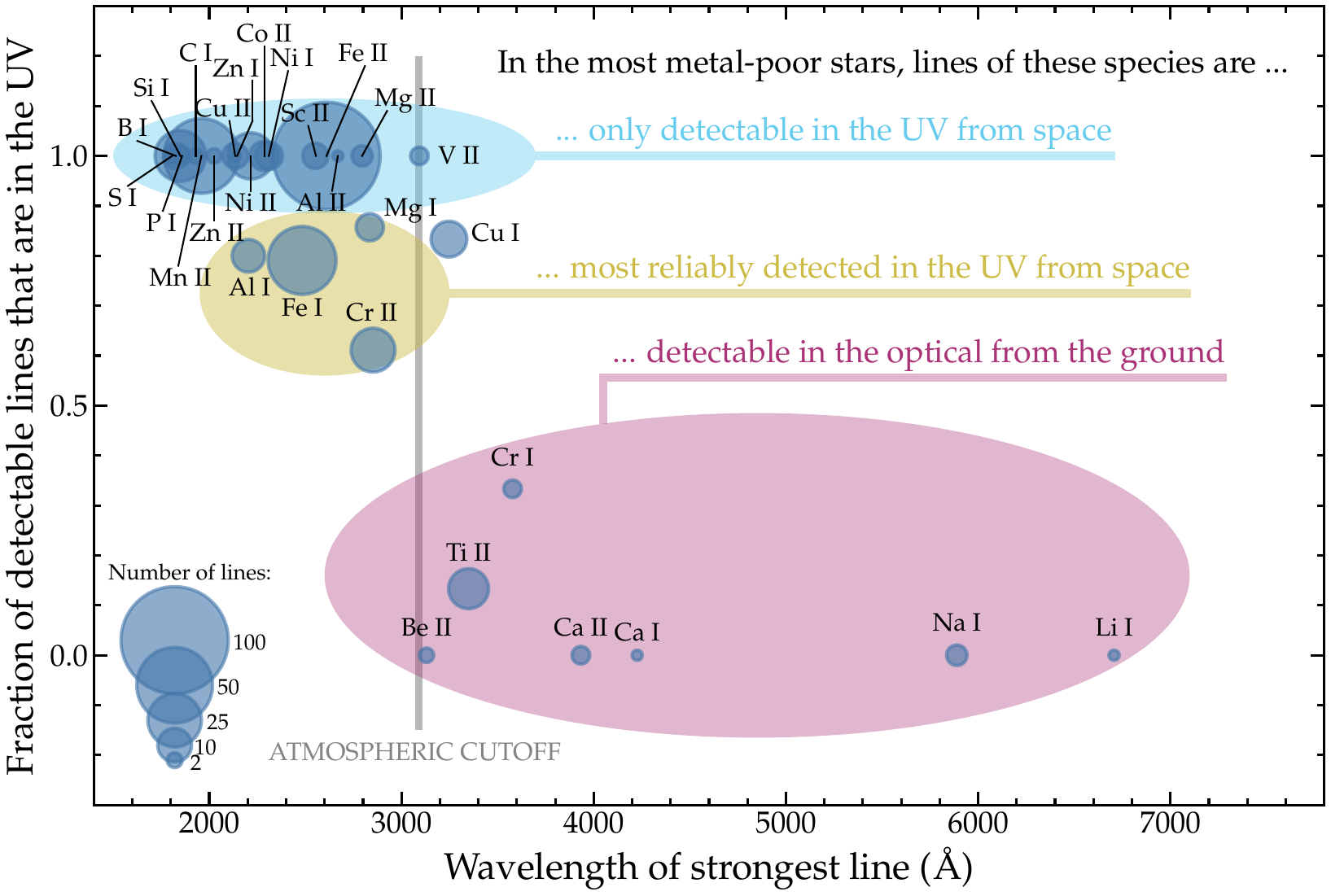}
\caption{\small Relationship between wavelength 
of the strongest absorption line of each metal species 
and the fraction of lines that could be detectable 
in the most metal-poor stars that are found in the UV.~
These data are mostly drawn from the 
National Institute of Standards and Technology (NIST)
Atomic Spectra Database (ASD) \citep{kramida24}
and references listed in the text.
\label{fig:wavel}
}
\end{center}
\end{figure*}

\textit{Detect one or more low-mass first stars that survive 
at the present day, 
demonstrating their existence 
or placing stronger constraints on their absence.}
The strategy for this objective is to derive upper limits 
on abundances of metals from the non-detection 
of UV lines in their spectra. 
The UV is necessary because most metals expected to be present 
in second- and subsequent-generation (``Pop~II'') stars,
and therefore absent in Pop~III stars, 
produce their strongest transitions in the UV spectral domain. 
Without those UV lines, the upper limits on abundances derived 
from the non-detection of optical or NIR lines 
are insufficient to exclude the presence of metals in these stars
\citep{keller14}.
As shown in Figure~\ref{fig:wavel}, 
key metals with strong UV transitions include 
C, Mg, Al, Si, P, S, Sc, V, Cr, Mn, Fe, Co, Ni, Cu, and Zn
\citep{morton03}.
Key metals with stronger optical lines than UV lines 
are limited to Na, Ca, and Ti.
This approach complements studies that aim to 
directly image highly magnified Pop III stars at high redshift
\citep{windhorst18,welch22}.

\textit{Derive abundances of metals in the low-mass 
long-lived second generation of stars.}
The strategy for this objective 
is to observe the most metal-poor stars known, 
which have been previously identified based on the 
weakness or absence of strong optical lines, 
and derive abundances or upper limits on abundances 
based on the strong UV transitions of metals found in their spectra
\citep{roederer16d}.
The UV is required for the same reasons described above, 
and the key metals with strong UV transitions are also 
the same ones listed above. 
Model constraints improve as more elements are detected 
\citep{tominaga14,salvadori19,koutsouridou23}.
This approach complements high-resolution spectroscopic 
studies of metals in the most metal-poor damped Lyman-$\alpha$ systems
\citep{cooke11,becker12,welsh23}.

\section{Physical Parameters}

\begin{figure*}[ht]
\begin{center}
\includegraphics[width=0.72\textwidth]{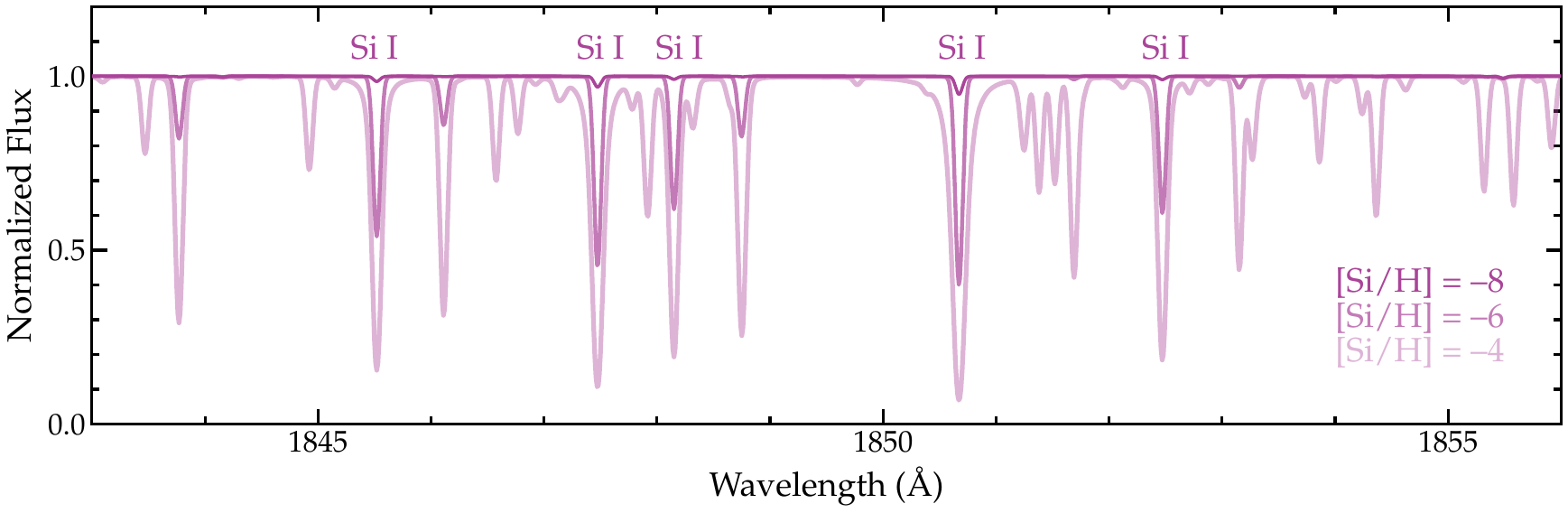}
\includegraphics[width=0.72\textwidth]{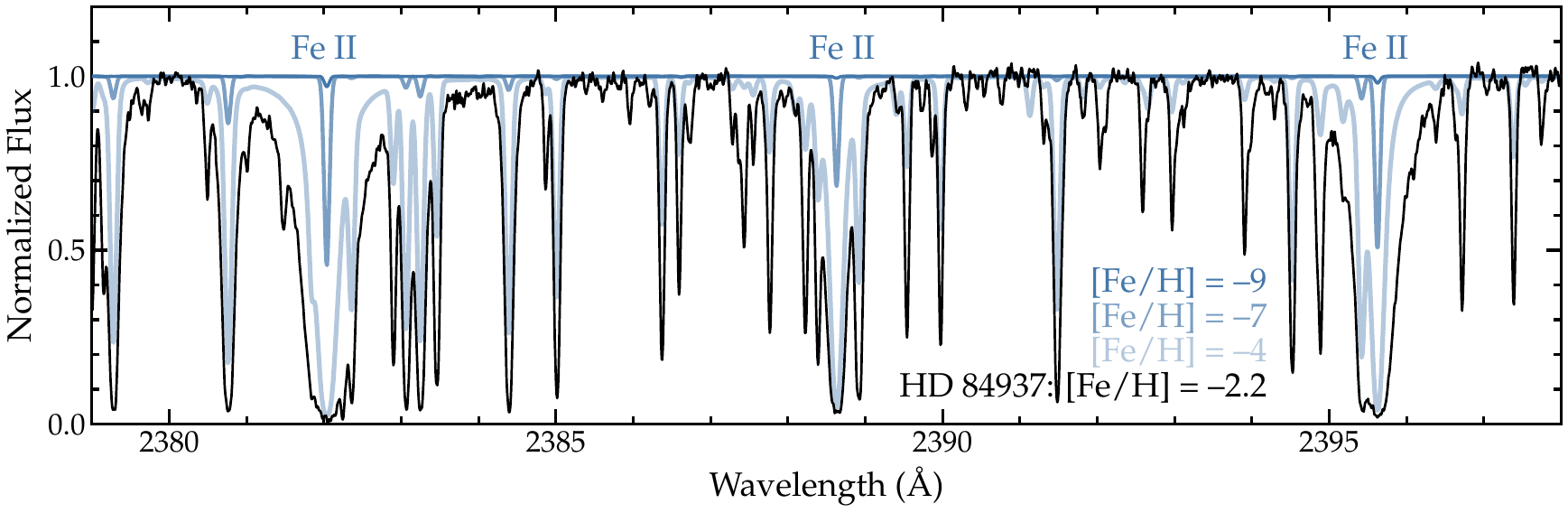}
\includegraphics[width=0.72\textwidth]{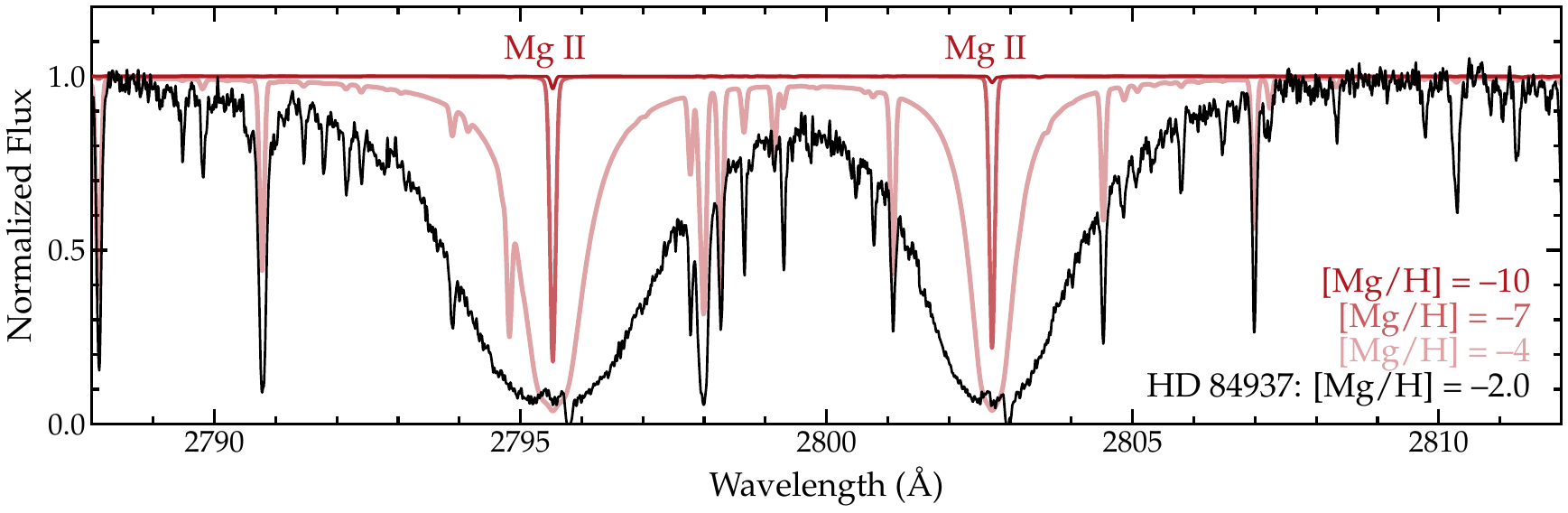}
\caption{\small 
Simulated UV spectra covering strong lines of 
Si~\textsc{i} (top), Fe~\textsc{ii} (middle), and Mg~\textsc{ii} (bottom).  
These spectra were generated using MOOG \citep{sneden73,sobeck11}
with 1D LTE models interpolated from the ATLAS9 grid \citep{castelli04}
for a typical G-type subgiant star with $T_{\rm eff}$ = 5500~K and 
$\log g$ = 3.5 and smoothed to $R\sim$~40,000.
Abundances of other metals have been scaled down proportionally 
with the element of interest.
The black line (middle and bottom panels) represents the spectrum 
of the bright, F-type metal-poor star \mbox{HD~84937} 
collected with HST/STIS/E230H \citep{peterson17}.
Lines of Si~\textsc{i}, Fe~\textsc{ii}, and Mg~\textsc{ii}
are still detectable 
in simulated spectra with S/N = 100 for the lowest abundances shown here, 
and upper limits based on non-detections would reach even lower abundances.
\label{fig:specplot}
}
\end{center}
\end{figure*}

This science case requires the detection of
UV metal absorption lines in FGK-type stars 
(i.e., in a second-generation star), 
or strong upper limits on the metal abundances 
derived from non-detection of these lines (i.e., in a first-generation star).

The second-generation descendants of the first stars 
have very low metal abundances \citep{ezzeddine17}, 
often [Fe/H] $< -4$, or 1/10,000$^{\rm th}$ the Solar iron abundance\footnote{%
[Fe/H] $\equiv \log_{10}(N_{\rm Fe}/N_{\rm H})_{\star} - \log_{10}(N_{\rm Fe}/N_{\rm H})_{\odot}$}.
Only a few tens of absorption lines are commonly found 
in the optical and near-infrared ($\lambda > 3100$~\AA) 
spectra of these stars, so only $\sim$~5--10 elements are regularly detected
(e.g., \citealt{aoki06he,caffau12,aguado18a}).
This situation limits the utility of these stars 
for understanding the nature of the first stars and first supernovae. 
Many other elements are expected to be present 
but are rarely detected. 
These elements constrain different aspects of supernova physics. 
For example, Co, Ni, and Zn provide the best constraints 
on the first stars' explosion energies and geometry \citep{ezzeddine19}.


Each element detected improves the model constraints, 
although the improvement varies 
depending on the various combinations of elements observed 
and the properties of each particular supernova model \citep{placco21}.
The strongest transitions of these elements are in the UV, 
below the atmospheric cutoff 
(e.g., 
Be~\textsc{i} 2348~\AA; 
B~\textsc{i} 1825, 2088, 2496~\AA; 
Mg~\textsc{ii} 2795, 2802~\AA;
Si~\textsc{i} 1850, 2124~\AA; 
P~\textsc{i} 1859, 2136~\AA; 
S~\textsc{i} 1807~\AA; 
Sc~\textsc{ii} 2555~\AA; 
V~\textsc{ii} 2683~\AA; 
Cr~\textsc{ii} 2055~\AA; 
Mn~\textsc{ii} 2605~\AA; 
Fe~\textsc{ii} 2343, 2382, 2395, 2404, 2585, 2598, 2599, 2607, 2611~\AA;
Co~\textsc{ii} 2286, 2580~\AA; 
Ni~\textsc{ii} 2165, 2216~\AA; 
Zn~\textsc{ii} 2062~\AA), 
requiring a high-resolution UV spectrograph in space for detection. 
Simulated $R \equiv \lambda/\Delta\lambda$ = 40,000 spectra 
of G-type metal-poor subgiant stars 
with S/N = 100 indicate that 3$\sigma$ upper limits 
based on non-detection of lines of
Si~\textsc{i} at 1850~\AA, 
Fe~\textsc{ii} at 2382~\AA, and
Mg~\textsc{ii} at 2795~\AA\
could reach 
[Mg/H] $< -10$, [Si/H] $< -8$, and [Fe/H] $< -9$
(Figure~\ref{fig:specplot}).

\begin{figure}[ht]
\begin{center}
\includegraphics[width=0.45\textwidth]{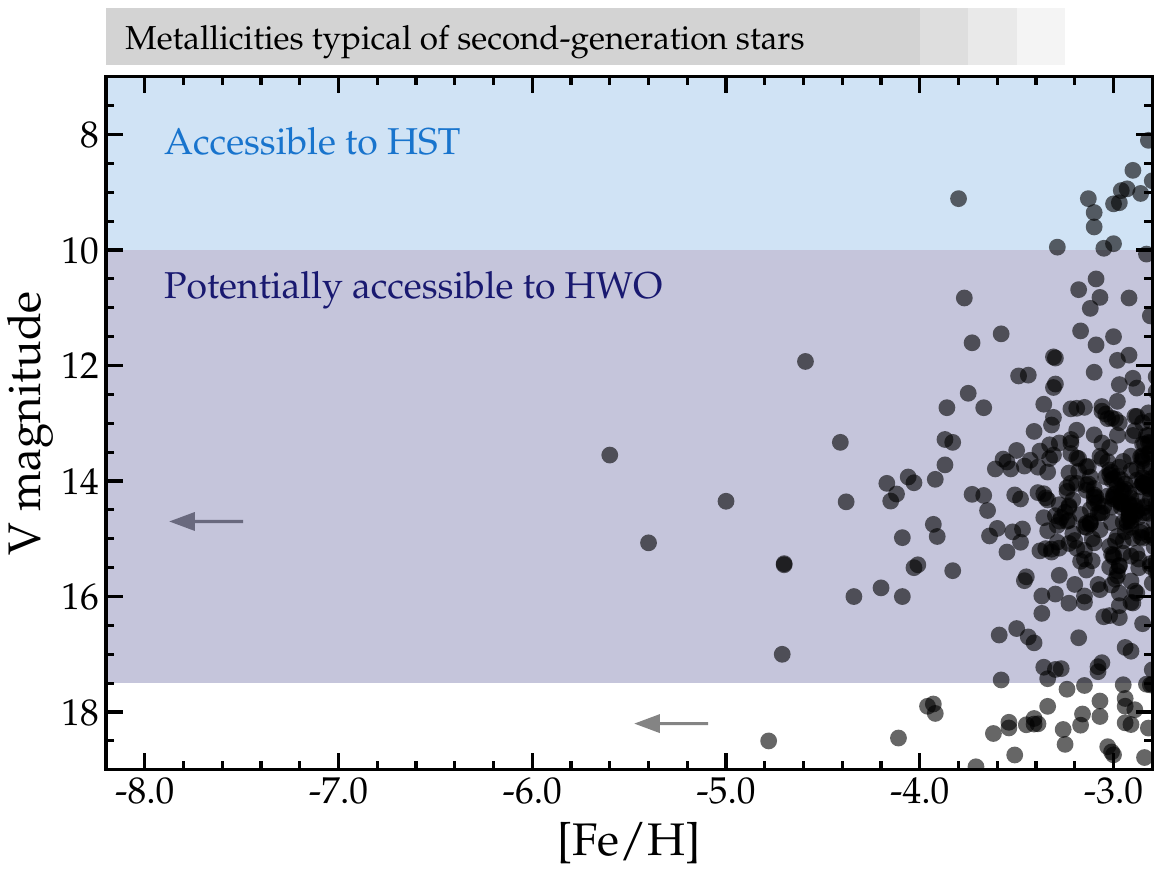}
\caption{\small 
Comparison of the metallicities of the most metal-poor stars 
and their accessibility to high-resolution UV spectroscopy with HST/STIS 
(blue band, top) and HWO (purple band, bottom).
Several dozen stars with [Fe/H] $< -4$, including several with 
only upper limits determined 
from optical spectra (left-pointing arrows), would 
be potentially observable with HWO.
\label{fig:vmag}
}
\end{center}
\end{figure}

Most stars known at present with [Fe/H] $< -4$ are found in the
Milky Way halo or its population of dwarf galaxies.
These stars are relatively faint, with $V >$~12
(Figure~\ref{fig:vmag}).
With the Hubble Space Telescope (HST), the current state of the art, 
we have been limited to studying only the brightest stars. 
Only one star with $V < 10$ and [Fe/H] $< -3.5$ is known at present 
(\mbox{BD~$+$44$^{\circ}$493}), and it has been observed
with the Space Telescope Imaging Spectrograph (STIS)
\citep{placco14bdp44}.
Only one star with [Fe/H] $< -4$ has been successfully observed 
with the Cosmic Origins Spectrograph (COS) 
(\mbox{HE~1327$-$2326}; \citealt{ezzeddine18}),
although unsuccessful efforts were made 
to observe a second one \citep{roederer17smss}.
HWO would overcome this fundamental limitation of HST 
and enable us to study much fainter stars for the first time.

Each star observed today reflects the yields of an individual supernova 
from the early Universe, so a larger sample of stars (Table~1)
increases the sample of 
(unobserved, but studied through their elemental yields) 
first-star supernovae by a direct 1-to-1 factor. 
This would revolutionize our understanding of the first stars, 
the first supernovae, and the first metals in the Universe.

\section{Description of Observations}

High spectral resolving power 
and high S/N ratios 
are essential to detect and accurately measure 
the relatively weak absorption lines produced by these elements
(Table~2).
Spectral coverage from 1800--3100~\AA\ 
should be attainable in a minimal number of exposures.
These values are based on experience with STIS or COS spectra.
The larger primary mirror of HWO and much greater instrument throughput 
(relative to STIS) enable similar observations to be made 
of much fainter stars. 
These fainter magnitudes
enable many more candidate second-generation 
(or first-generation Pop~III stars, 
should such candidates be identified in the future) 
to be observable with HWO.

The spectral range from 1400--1800~\AA\ 
is unexplored in metal-poor stars. 
This spectral region may enable detection of absorption lines 
of previously undetected elements, 
but that possibility is unconfirmed at present.
This spectral region is listed 
in the ``breakthrough'' column in Table~2,
but otherwise 1800--3100~\AA\ should generally be sufficient.
This full wavelength range should be attainable in relatively few setups
to maximize the observing efficiency.

The field-of-view (FOV) and angular resolution 
are not constraints that will drive this science case.
The FOV only needs to be large enough to observe one 
point-source target star at a time on a narrow slit.
Target stars are unlikely to be in crowded fields 
(within $\approx$~0\farcs1 or so) 
with other objects of comparable brightness 
(within $\sim$~5~mag or so). 
At 2000~\AA, the diffraction limit 
(1.22$\lambda$/$D$) for a 6.5~m telescope is $\approx$~0\farcs008, 
far smaller than angular separation to any potential neighbor objects.

Target stars will likely be spread across the entire sky,
and in few if any cases will they be within a few arcseconds of each other, 
so the number of fields is equivalent to the number of targets.

{\bf Acknowledgements.}
We acknowledge funding from the US National Science Foundation 
(AST~2205847 to IUR, AST-2206263 to RE)
and numerous grants 
(including GO-15657 and GO-15951 to IUR and RE)
provided by NASA through the Space Telescope
Science Institute, which is operated by the Association of 
Universities for Research in Astronomy, Incorporated, under NASA
contract NAS5-26555.
We thank the NIST ASD team; 
this work would not have been possible without 
their efforts to maintain this database over many decades.

\begin{table*}[!ht]
\caption{Physical Parameters}
\smallskip
\begin{center}
{\small
\begin{tabular}{ccccc}  
\tableline
\noalign{\smallskip}
\textbf{Physical Parameter} &
\textbf{State of the Art} &
\textbf{Incremental Progress} &
\textbf{Substantial Progress} &
\textbf{Major Progress} \\
 & & \textbf{(Enhancing)} & \textbf{(Enabling)} & \textbf{(Breakthrough)} \\
\noalign{\smallskip}
\tableline
\noalign{\smallskip}
Number of stars with {[Fe/H]} $< -4$ & 1 observed in the UV & 3 & 10 & 50 \\
\noalign{\smallskip}
Number of elements per star  & 5--10, including optical detections & 10 & 15 & 20 \\
\noalign{\smallskip}
\tableline\
\end{tabular}
}
\end{center}
\end{table*}

\begin{table*}[!ht]
\caption{Observational Requirements}
\smallskip
\begin{center}
{\small
\begin{tabular}{ccccc}  
\tableline
\noalign{\smallskip}
\textbf{Observational Requirement} & 
\textbf{State of the Art} & 
\textbf{Incremental Progress} & 
\textbf{Substantial Progress} & 
\textbf{Major Progress} \\
 & & \textbf{(Enhancing)} & \textbf{(Enabling)} & \textbf{(Breakthrough)} \\
\noalign{\smallskip}
\tableline
\noalign{\smallskip}
$V$ magnitude of star with {[Fe/H]} $< -4$ & 10 & 12 & 14 & 18 \\
\noalign{\smallskip}
Spectroscopic resolving power & 30,000 & 60,000 & 100,000 & 100,000 \\
\noalign{\smallskip}
Wavelength range (\AA) & 
1800--3100 & 1800--3100 & 1800--3100 & 1400-3100 \\
\noalign{\smallskip}
S/N after co-adds & 20 & 50 & 100 & 100 \\
\noalign{\smallskip}
FOV & (one point source) & (one point source) & (one point source) & (one point source) \\
\noalign{\smallskip}
Angular resolution at 2000~\AA\ & $<$ 0\farcs1 & $<$ 0\farcs1 & $<$ 0\farcs1 & $<$ 0\farcs1 \\
\noalign{\smallskip}
Number of fields & 1 & 3 & 10 & 50 \\
\noalign{\smallskip}
\tableline\
\end{tabular}
}
\end{center}
\end{table*}
\noindent

\bibliography{roederer_firststars}

\end{document}